\documentclass[twocolumn,prl,aps,showpacs,superscriptaddress]{revtex4-1}
\usepackage{graphicx,graphics}
\usepackage{amsmath}
\usepackage{amssymb}
\usepackage{epstopdf}
\usepackage{amstext}
\usepackage{amsthm}
\usepackage{amsfonts}
\usepackage{latexsym}
\usepackage{array}
\usepackage{xfrac}
\usepackage{lipsum}
\usepackage[toc,page]{appendix}
\usepackage{subfigure}
\newcommand{\igbjd}[1]{}\newcommand{\beqa}{\begin{eqnarray}}
\newcommand{\eeqa}{\end{eqnarray}}
\newcommand{\beq}{\begin{equation}}
\newcommand{\eeq}{\end{equation}}
\usepackage[normalem]{ulem}

\usepackage[colorlinks=true,citecolor=Cerulean,linkcolor=RubineRed,urlcolor=Cerulean,pdftex]{hyperref}

\graphicspath{{figures/}}
\usepackage{lineno}
\usepackage{xcolor}
\definecolor{Cerulean}{rgb}{0.,0.59,0.835}
\definecolor{RubineRed}{rgb}{0.61,0.07,0.12}
\usepackage[export]{adjustbox}

\newcommand{\1}{{\text A}}
\newcommand{\2}{{\text B}}

\begin{document}
\title{Quantum halo states in two-dimensional dipolar clusters}
\author{G. Guijarro}
\author{G. E. Astrakharchik}
\author{J. Boronat}
\affiliation{Departament de F\'isica, Campus Nord B4-B5, Universitat Polit\`ecnica de Catalunya, E-08034 Barcelona, Spain}
%\date{\today}
\date{January 15, 2021}
\begin{abstract}
A halo is an intrinsically quantum object defined as a bound state of a spatial size which extends deeply
into the classically forbidden region. Previously, halos have been observed in bound states of two and
less frequently of three atoms. Here, we propose a realization of halo states containing as many as six
atoms. We report the binding energies, pair correlation functions, spatial distributions, and sizes of
few-body clusters composed by bosonic dipolar atoms in a bilayer geometry. We find two very distinct halo
structures, for large interlayer separation the halo structure is roughly symmetric and we discover an
unusual highly anisotropic shape of halo states close to the unbinding threshold. Our results open
avenues of using ultracold gases for the experimental realization of halos with the largest number of
atoms ever predicted before. 
\end{abstract}
\maketitle

\section{Introduction}
One of the most remarkable aspects of ultracold quantum gases is their versatility, which permits to
bring ideas from other areas of physics and implement them in a clean and highly controllable manner.
Some of the examples of fruitful interdisciplinary borrowings include Efimov states, originally
introduced in nuclear physics and observed in alkali atoms~\cite{EFIMOV1970563,kraemer2006evidence,Naidon_2017},
lattices created with counter-propagating laser beams~\cite{sachdev_2009,Bloch2005} as models for crystals in
condensed matter physics, and Bardeen-Cooper Schrieffer (BCS) pairing theory first introduced to explain
superconductivity and later used to describe two-component Fermi gases~\cite{ReviewFermiGases,Zwerger2012book}.
In the present work, we exploit the tunability of ultracold gases to demonstrate the existence of halo states
with a number of atoms never predicted before. Originating in nuclear
physics~\cite{RevModPhys.66.1105,TANIHATA1985,PhysRevLett.55.2676}, halo dimer states have been studied and
experimentally observed in ultracold gases~\cite{RevModPhys.82.1225}.

A halo is an intrinsically quantum object and it is defined as a bound state with a wave function that
extends deeply into the classically forbidden region~\cite{Jensen2004,Riisager_2013}. These states are
characterized by two simultaneous features: a large spatial size and a binding energy which is much
smaller than the typical energy of the interaction. One of the most dramatic examples of a halo system,
experimentally known, is the Helium dimer ($^4\mathrm{He}_2$), which is about ten times more extended
than the size of a typical diatomic molecule~\cite{RevModPhys.82.1225}.

While most of the theoretical and experimental studies of halos have been carried in three dimensions
(3D)~\cite{RIISAGER1992393,PhysRevC.49.201,PhysRevLett.113.253401,Stipanovi2017QuantumHS,Stipanovic2019,
PhysRevLett.103.033004,PhysRevA.90.043631,Kievsky_2014}, there is an increasing interest in halos in two
dimensions (2D)~\cite{Nielsen1997,Nielsen1999,Jensen2004}. In fact, two dimensions are especially
interesting as halos in 2D have different properties~\cite{Jensen2004} of the 3D ones. A crucial
difference between 3D and 2D geometries is that lower dimensionality dramatically enhances the
possibility of forming bound states. If the integral of the interaction potential $V(\rho)$ over all the
space is finite and negative, $V_{k=0} = \int V({\bf {\rho}})d{\bf \rho}<0$, this is always sufficient
to create a two-body bound state in 2D but not necessarily in 3D. Furthermore, the energy of the
bound-state is exponentially small in 2D and it can be expressed as 
$E = -\hbar^2/(2ma^2) \exp[-\hbar^2|V_{k=0}|/(2\pi m)]$~\cite{LandauLifshitz_iii}, where $a$ is the
typical size of the bound state. An intriguing possibility arises when such integral is exactly equal to
zero, $V_{k=0} = 0$, as a priori it is not clear if a bound state exists. This situation exactly happens
in a dipolar bilayer in which atoms or molecules are confined to two layers separated by a distance $h$
and the dipolar moments $d$ are aligned perpendicular to the plane of motion by an external field. The
interaction between atoms of different layers is given by $V(\rho) = d^2(\rho^2-2h^2)/(\rho^2+h^2)^{5/2}$,
where $\rho$ is the in-plane distance. The vanishing Born integral has first lead to conclusions that the
two-body bound state disappears when the distance between the layers is large~\cite{Yudson1997} although
later it was concluded that the bound state exists for any
separation~\cite{Shih2009,Armstrong2010,Klawunn2010,Baranov2011,Volosniev2011,Macia2014}, consistently
with Ref.~\cite{Simon1976}. A peculiarity of this system is that the bound state is extremely weakly
bound in $h\to\infty$ limit. That is, a potential with depth $V(\rho=0) = -d^2/h^3$ and width $h$ would
be expected to have binding energy equal to $E = -\hbar^2/(2ma^2)\exp(-\text{const}\cdot r_0/h)$ where
$r_0 = md^2/\hbar^2$ is the characteristic distance associated with the dipolar interaction and $m$ is
the particle mass. Instead, the correct binding energy, 
$E = - 4\hbar^2/mh^2\exp(-8r_0^2/h^2 + O(r_0/h))$~\cite{Klawunn2010,Baranov2011} is much smaller as
it has $h^{-2}$ in the exponent and not the usual $h^{-1}$. This suggests that the bilayer configuration
is very promising for the creation of a two-body halo state. Moreover, the peculiarity of the bilayer
problem has resulted in the controversial claim that the three- and four-body~\cite{Volosniev2012} bound
states never exist in this system, and only very recently it has been predicted that actually, they are
formed~\cite{PhysRevA.101.041602}. 

In the present work, we analyze the ground-state properties of few-body bound states of dipolar bosons
within a two-dimensional bilayer setup, as candidates for halo states. In particular, we study the
ground-state of up to six particles occupying the A and B layers, with A and B denoting particles in
different planes. In order to find the exact system properties, we rely on the diffusion Monte Carlo
(DMC) method~\cite{BoronatCasulleras1994} with pure estimators~\cite{CasullerasBoronat1995}, which has
been used previously to get an accurate description of quantum halo states in Helium
dimers~\cite{PhysRevLett.113.253401}, trimers and tetramers~\cite{Stipanovi2017QuantumHS,Stipanovic2019}.
In addition, we report relevant structure properties of the clusters, such as the spatial density
distributions and the pair distribution functions for characteristic interlayer separations.

\section{Hamiltonian}
We consider two-dimensional systems consisting from two to six dipolar bosons of mass $m$ and dipole moment
$d$ confined to a bilayer setup. All the dipole moments are oriented perpendicularly to the layers making
the system always stable. In this configuration the angular dependence of the dipolar interaction vanishes.
In our model, we suppose that the confinement to each plane is so tight that there is no interlayer
tunneling and excitations into the excited levels of the tight confinement are suppressed.
The Hamiltonian of this system is
\begin{equation}
\begin{aligned}
H=&-\frac{\hbar^2}{2m}\sum_{i=1}^{N_\1}\nabla^2_i-\frac{\hbar^2}{2m}
\sum_{\alpha=1}^{N_\2}\nabla_\alpha^2\\
+&\sum_{i<j}\frac{d^2}{\rho^3_{ij}}+\sum_{\alpha<\beta}\frac{d^2}
{\rho^3_{\alpha\beta}}+\sum_{i\alpha}\frac{d^2(\rho_{i\alpha}^2-2h^2)}
{(\rho_{i\alpha}^2+h^2)^{5/2}}\;,
\end{aligned}
\label{Hamiltonian}
\end{equation}
where $h$ is the distance between the layers.
The terms in the first row of the Hamiltonian~(\ref{Hamiltonian}) are the kinetic energy of $N_\1$ dipoles
in the bottom layer and $N_\2$ dipoles in the top layer; the first two terms in the second row correspond
to the intralayer dipolar interactions of $N_\1$ and $N_\2$ bosons; and the last one accounts for the
interlayer interactions. The in-plane distance between two bosons belonging to the same layer is denoted
by $\rho_{ij(\alpha\beta)}=|{\bf{\rho}}_{i(\alpha)}-{\bf{\rho}}_{j(\beta)}|$, and belonging to different
layers by $\rho_{i\alpha}=|{\bf{\rho}}_{i}-{\bf{\rho}}_{\alpha}|$, where $\rho_i$ is the in-plane position.
We use the characteristic dipolar length $r_0=md^2/\hbar^2$ and energy $E_0=\hbar^2/(mr_0^2)$ as units of
length and energy, respectively. We use $\rho$ for 2D in-plane distances and $r$ for 3D distances.

Dipoles in the same layer are repulsive, with an interaction decaying as $1/\rho^3$. However, for dipoles
in different layers the interaction is attractive for small in-plane distance $\rho$ and repulsive for
larger $\rho$. In other terms, a dipole in the bottom layer induces attractive and repulsive zones for
a dipole in the top layer. Importantly, the area of the attractive cone increases with the distance
between layers $h$, making the formation of few-body bound states more efficient.

\section{Structure of the bound states}
We first analyze the structure of few-body clusters, composed by up four particles, as a function of the
interlayer separation $h$. To this end, we calculate the pair distribution function $g_{\sigma\sigma^{'}}(r)$,
which is proportional to the probability of finding two particles at a relative distance $r$. In the case
of the ABB trimers and AABB tetramers, we also determine the ground-state density distributions for different
values of the interlayer separation.

\subsection{AB Dimer}
\begin{figure}
\centering
\includegraphics[width=0.48\textwidth]{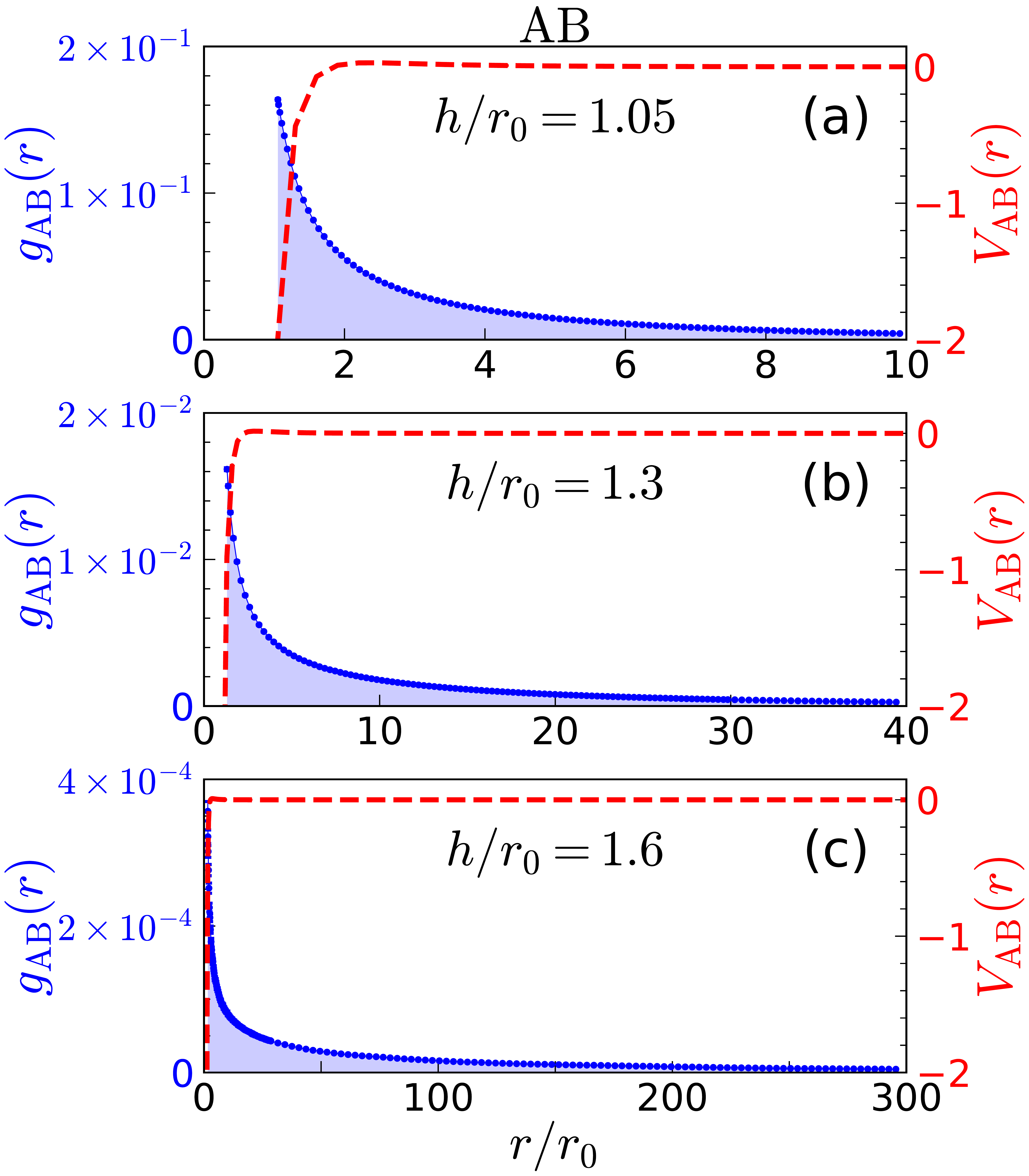}
\caption{Interlayer pair distributions $g_{\1\2}(r)$ (left-axis, blue curves) and dipolar potentials
$V_{\1\2}(r)$ (right-axis, red curves) for AB and for three values of the interlayer distance
$h/r_0=1.05$, $1.3$ and $1.6$. Notice the different scales in the $r$ axis.}
\label{Fig:grAB}
\end{figure}

The AB dimer is strongly bound for $h\lesssim r_0$ and its energy decays exponentially in the limit of
large interlayer separation~\cite{Baranov2011}. In order to understand how the cluster size changes
with $h/r_0$, we show in Fig.~\ref{Fig:grAB} the interlayer pair distributions $g_{\1\2}(r)$
(left-axis, blue curves) for three values of $h/r_0$. The strong-correlation peak of $g_{\1\2}$ at
$h/r_0$ is due to the interlayer attraction $V_{\1\2}(r)$ at short distances, also shown in the
right-axis of the figure (red curves). For the cases shown in Fig.~\ref{Fig:grAB} we notice that
$g_{\1\2}$ are very wide in comparison to the interlayer distance $h/r_0$ reflecting the exponential
decay of the bound state. The tail at large distances becomes longer as the interlayer distance increases. 

\subsection{ABB Trimer}
\begin{figure*}[htp]
    \centering
    \subfigure{\includegraphics[width=0.48\textwidth]{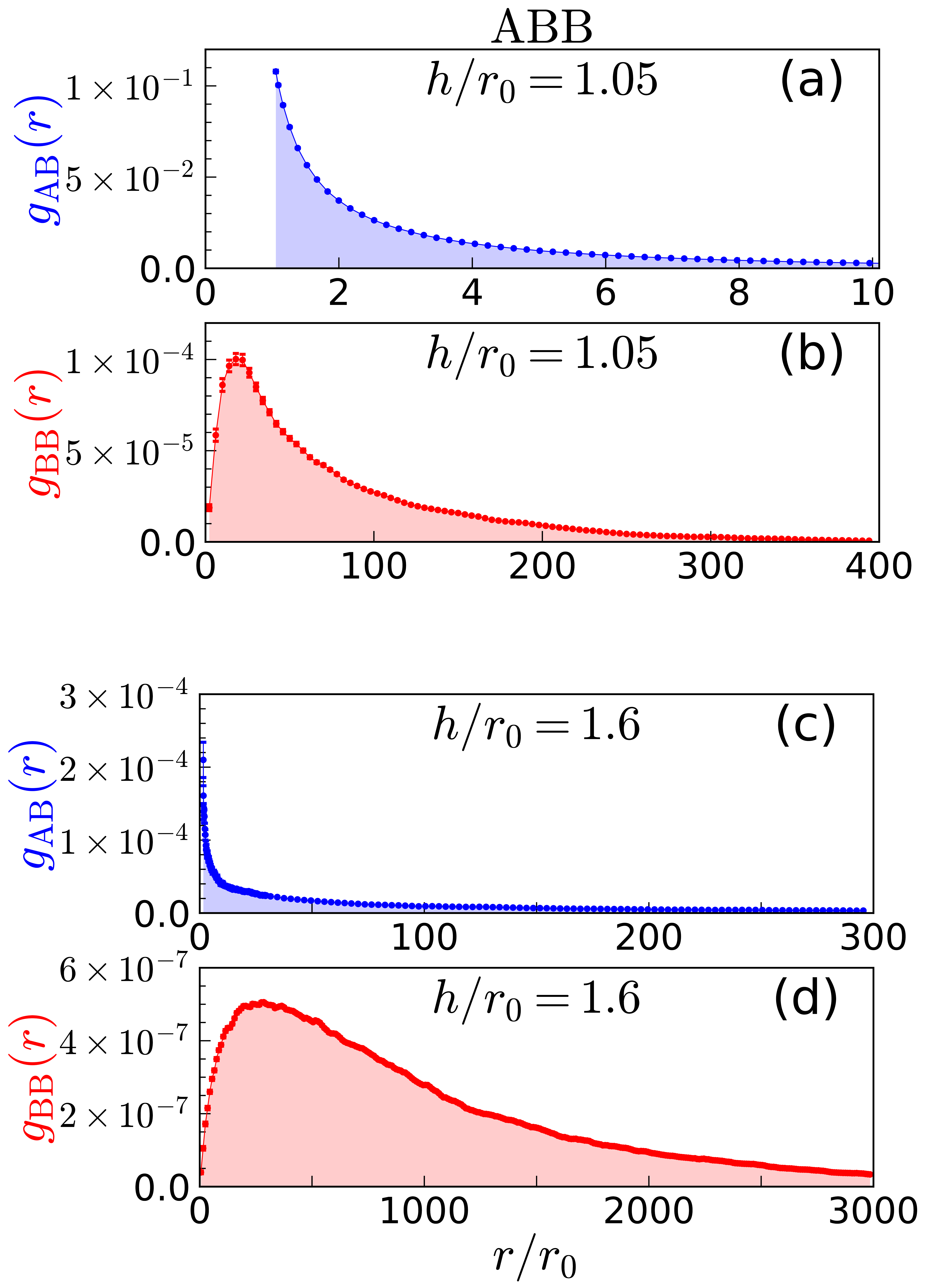}}\quad\quad
    \subfigure{\includegraphics[width=0.48\textwidth]{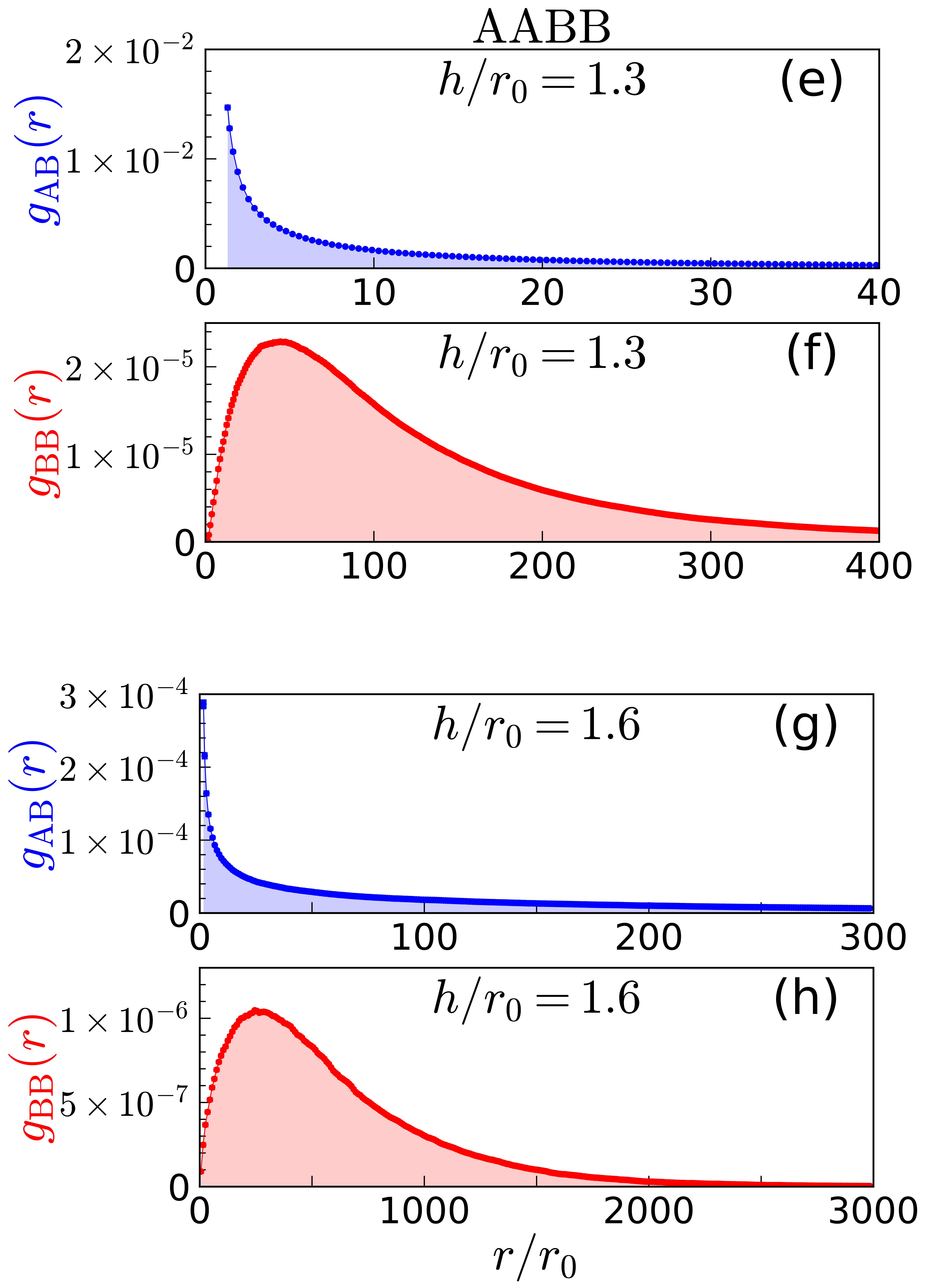}}
    \caption{Interlayer and intralayer pair distributions, $g_{\1\2}(r)$
         and $g_{\2\2}(r)$, for ABB (a, b, c, d) and AABB (e, f, g, h) 
         clusters, and for different values of
         the interlayer distance $h/r_0$.}  
	\label{Fig:PairDistributions}
\end{figure*}

The ABB trimer is bound for large enough separation between the layers $h/r_0>0.8$ while, for smaller
separations, it breaks into a dimer and an isolated atom ~\cite{PhysRevA.101.041602}. The trimer
binding energy is vanishingly small for $h\approx h_c$, with $h_c\simeq 0.8 r_0$, and it  becomes
larger as $h$ is increased, reaching its maximum absolute value at $h/r_0\approx 1.05$. Then, it
vanishes again in the limit of $h\to\infty$~\cite{PhysRevA.101.041602}. We report the intralayer
and interlayer pair distributions, $g_{\2\2}(r)$ and $g_{\1\2}(r)$ respectively, in
Fig.~\ref{Fig:PairDistributions} for strongly-~(a,~b) and weakly-bound~(c,~d) trimers. We observe
that the $g_{\1\2}$ distributions are very wide in comparison to $h$, similarly to what has been
observed in Fig.~\ref{Fig:grAB} for dimers. The same-layer distribution $g_{\2\2}$ vanishes when
$r/r_0 \to 0$ as a consequence of the strongly repulsive dipolar intralayer potential at short
distances. As $r$ increases, $g_{\2\2}$ exhibits a maximum, next it monotonically decreases with
$r/r_0$. For a weakly-bound trimer ($h/r_0=1.6$), both $g_{\1\2}$ and $g_{\2\2}$ produce long tails
at large distances.

The trimer is weakly bound close to the threshold, $h\to h_c$, and for large interlayer separation,
$h\to\infty$, but its internal structure in those two limits is significantly different. This can
be seen in Fig.~\ref{Fig:SpatialDistributions}~(a,~b), where we plot the trimer ground-state
spatial distribution for $h/r_0=1.05$ and 1.6. The spatial distribution is shown as a function of
the distance between two dipoles in the same layer $|{\bf r}_1^{\2}-{\bf r}_2^{\2}|$
(horizontal axis) and the minimal distance between dipoles in different layers 
$\mathrm{min}\{|{\bf r}_1^{\1}-{\bf r}_1^{\2}|, |{\bf r}_1^{\1}-{\bf r}_2^{\2}|\}$ (vertical axis).
For large separation between layers, shown in Fig.~\ref{Fig:SpatialDistributions}(b) for $h/r_0=1.6$,
the distances between AB and BB atoms are all of the same order, revealing an approximately symmetric 
structure. However, by decreasing the distance between layers the particle distribution becomes
significantly asymmetric. For $h/r_0=1.05$ (\ref{Fig:SpatialDistributions}(a)), we observe that the
trimer spatial distribution is elongated: two dipoles in different layers are close to each other
while the third one is far away. Regardless of the interlayer separation, the pair $\1\2$ is, on
average, closer than the $\2\2$ pair. As the system approaches to the threshold value, the trimer
becomes more extended and breaks into a dimer and a single atom at $h/r_0\approx 0.8$.

\subsection{AABB Tetramer}
\begin{figure*}[htp]
  \centering
  \subfigure{\includegraphics[width=1.0\textwidth]{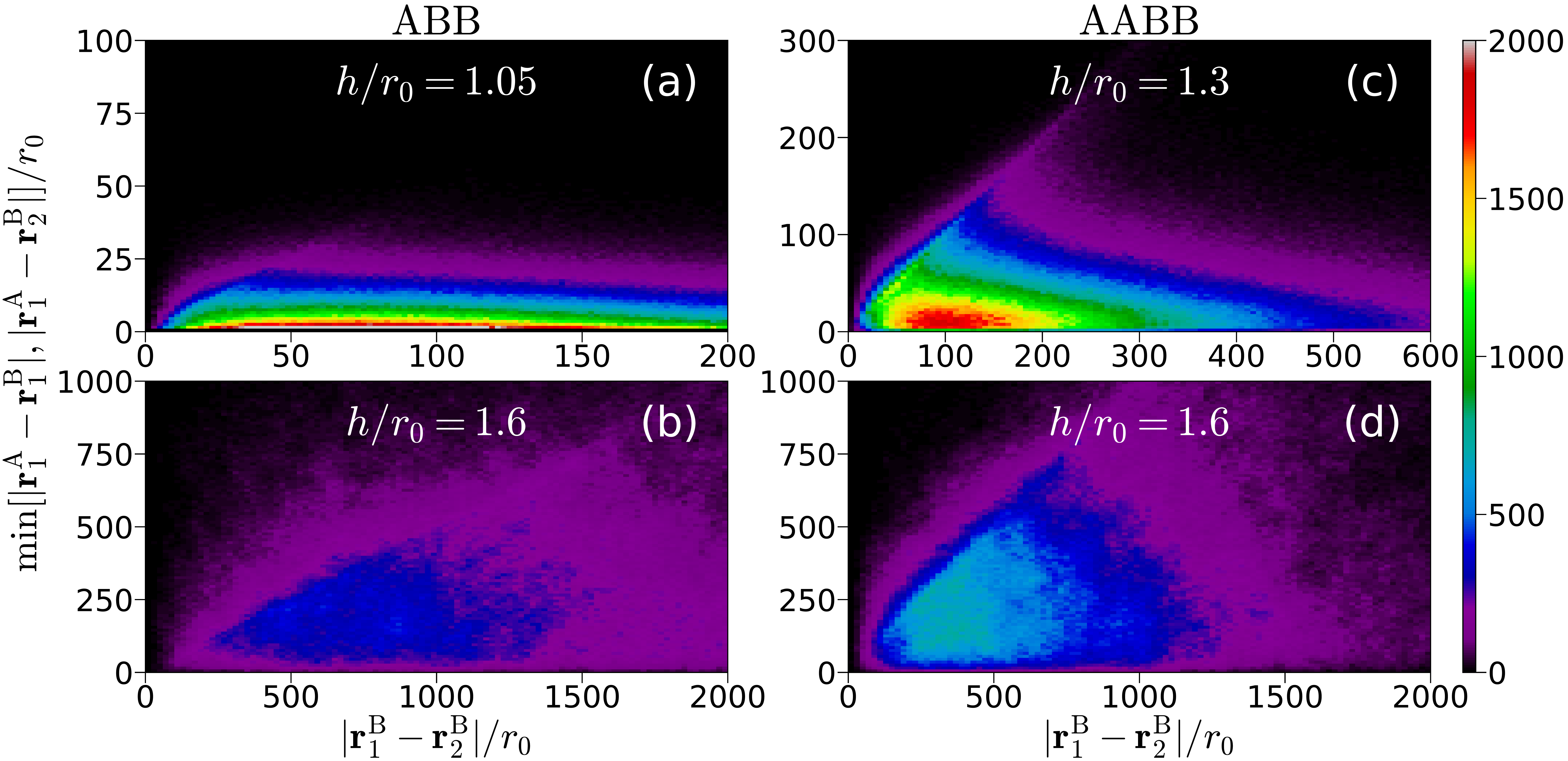}}
  \caption{Spatial structure of the ground-state for ABB trimer~(a,~b) and AABB tetramer~(c,~d) for
           different values of the interlayer distance. The distance between two dipoles in the same
           layer is plotted in the horizontal axis and the minimum distance between dipoles in
           different layers is shown in the vertical axis.}  
  \label{Fig:SpatialDistributions}
\end{figure*}

As we have shown in the previous Section, an ABB trimer dissolves into a dimer and an atom when
$h\approx h_c$. Here, we address the structure properties of the balanced case for a tetramer,
in which the number of A and B atoms is the same. The AABB tetramer is weakly bound for large
values of $h/r_0$. When the distance between layers decreases, the tetramer becomes unbound at
$h/r_0\approx1.1$ and splits into two AB dimers~\cite{PhysRevA.101.041602}.

The pair distributions $g_{\1\2}$ and $g_{\2\2}$ for AABB are shown in
Fig.~\ref{Fig:PairDistributions}(e, f, g, h) for two characteristic values of the interlayer
distance $h/r_0$. We observe a behavior that is similar to that previously reported for ABB.
That is, both $g_{\1\2}$ and $g_{\2\2}$ are compact for the deepest bound state
($h/r_0=1.3$ for AABB) and become diffuse, showing long tails at large distances, when it
turns to a weakly-bound state ($h/r_0=1.6$).   

The ground-state spatial distributions for the symmetric tetramer are shown in
Fig.~\ref{Fig:SpatialDistributions} (c, d). We observe that for large separation $h$, i.e.,
when the tetramer is weakly bound, it has large spatial extension and the distances between
AA and AB pairs are of the same order. As the interlayer separation is progressively decreased,
the tetramer size decreases and its structure becomes anisotropic. In this case, the distance
between dipoles in the same layer is several times larger than the distance between dipoles
in different layers. When the tetramer approaches to the threshold for unbinding the cluster
becomes even more elongated and it breaks into two AB dimers at $h/r_0\approx1.1$.

\section{Quantum halo characteristics}
\begin{figure*}[htp]
    \centering
    \includegraphics[width=0.9\textwidth]{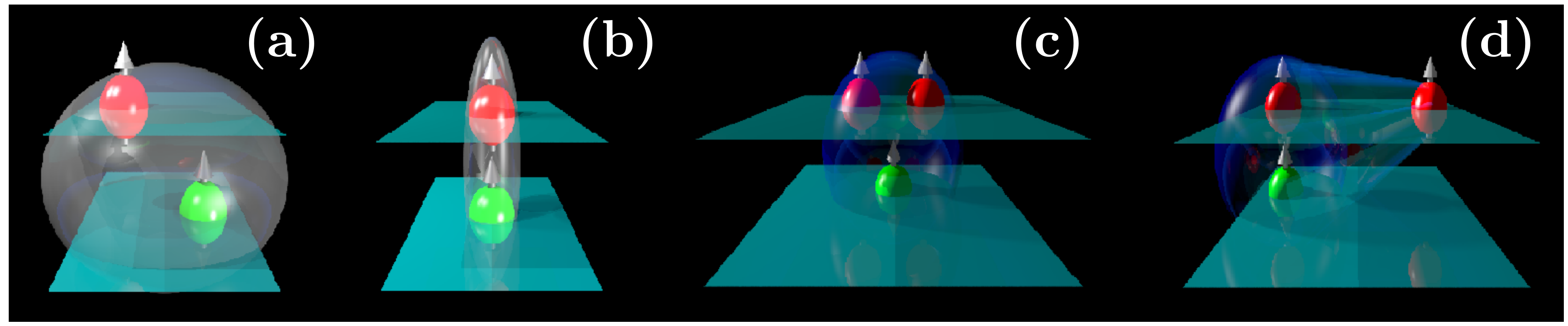}
    \includegraphics[width=0.49\textwidth]{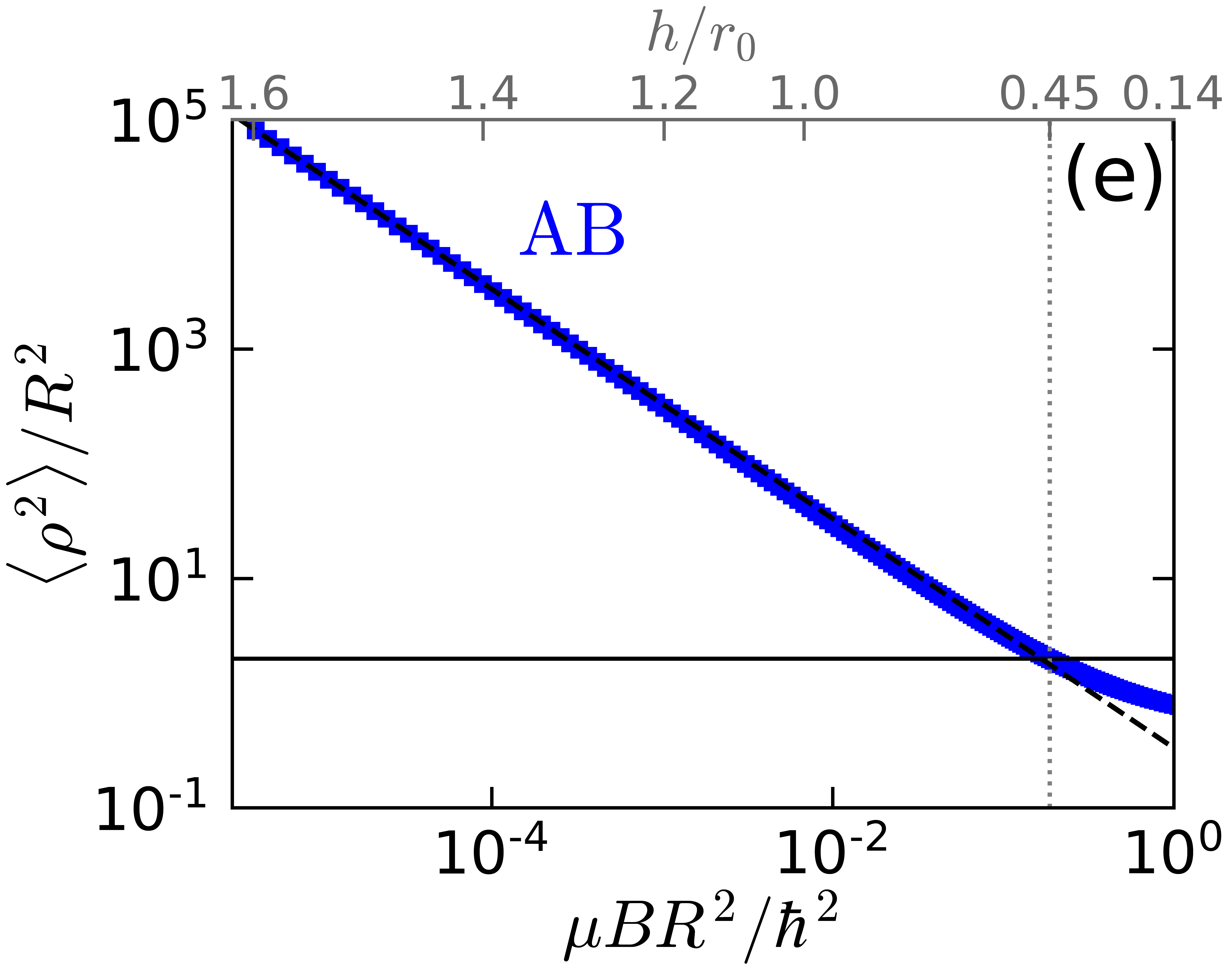}
    \includegraphics[width=0.49\textwidth]{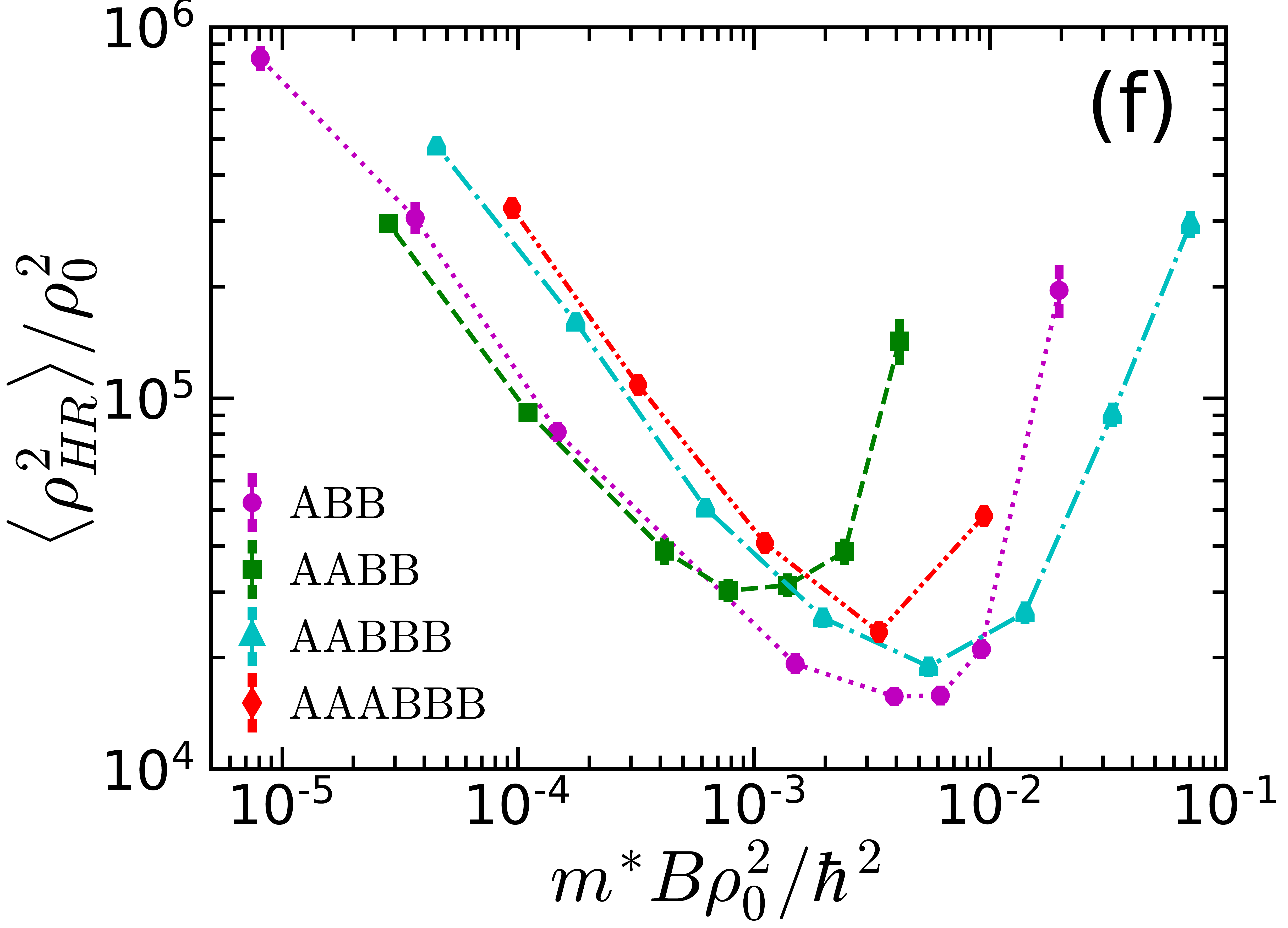}
    \caption{Top panel: Schematic representation of the AB and ABB states in two limits:
        (a) AB is a halo state; (b) AB is not a halo state; (c) ABB $h\to \infty$; (d) ABB $h\to h_c$.
        Bottom panel: (e) Size $\langle\rho^2\rangle/R^2$ vs ground-state energy $\mu BR^2/\hbar^2$
        scaling plot for two-body halos. The horizontal line is the quantum halo limit and the dashed
        one is $\langle \rho^2\rangle/R^2=\hbar^2/(3\mu BR^2)$, which is a zero-range approximation for
        two-body halos in two-dimensions~\cite{Jensen2004}. (f) Size $\langle\rho_{HR}^2\rangle/\rho_0^2$
        vs ground-state energy $m^*B\rho_0^2/\hbar^2$ scaling plot for three- up to six-body halos.}  
    \label{Fig:HaloStates}
\end{figure*}   

A halo is a quantum bound state in which particles have a high probability to be found in the
classically forbidden region, outside the range of the interaction potential. The key characteristics
of a halo are its extended size and binding energies much smaller than the typical energy of the
interaction. In order to classify a system as a halo state, one typically introduces two scaling
parameters with which the size and the energy are compared. The first parameter is the scaling length
$R$. For two-body systems one commonly chooses $R$ as the outer classical turning point. The second
parameter is the scaling energy $\mu BR^2/\hbar^2$, where $\mu$ is the reduced mass and $B$ is the
absolute value of the ground-state energy of the cluster. The size of a cluster is usually quantified
through its mean-square radius $\langle \rho^2 \rangle$, where $\rho$ is the interparticle distance.
A two-body quantum halo is then defined by the condition $\langle \rho^2 \rangle/R^2>2$, which means
that the system has a probability to be in the classically forbidden region larger than 50$\%$.

The dipolar interaction in the bilayer geometry has vanishing Born integral and thus, the AB dimer 
can show an enhancement of its halo properties. In Fig.~\ref{Fig:HaloStates} (e), we show the scaling
plot for the dipolar dimers, corresponding to interlayer distance  from $h/r_0=0.14$ to 1.6, as
indicated on the upper axis. All dimers which lie above the halo limit $\langle \rho^2\rangle/R^2=2$
(horizontal line in Fig.~\ref{Fig:HaloStates} (e)) are halo states and follow a universal scaling
law $\langle \rho^2\rangle/R^2=\hbar^2/(3\mu BR^2)$, shown with a dashed line in the
figure~\cite{Jensen2004}. This is exactly the case for all dimers with interlayer separations
$h/r_0 > 0.45$. This threshold value is close to the characteristic value, $h/r_0=0.5$, for which the
dimer binding energy is approximately equal to the typical energy of the dipolar interaction
$E_{\1\2} \approx \hbar^2/(mr_0^2)$.

While AB dimers exist for any interlayer separation, ABB trimers and AABB tetramers are self-bound
for large $h$ values, where AB dimers are in fact halo states. Thus, it can be anticipated that
these few-body bound states are also halos. The sizes of three- and four-body systems are measured
in terms of the mean-square hyperradius~\cite{Jensen2004},
\begin{equation}
    \rho_{HR}=\sqrt{\frac{1}{m^*M}\sum_{i<k}m_im_k({\boldsymbol{\rho}}_i-{\boldsymbol{\rho}}_k)^2}\;,
\end{equation}
where $m^*$ is an arbitrary mass unit, $M$ is the total mass of the system, and $m_i$ is the mass
of particle $i$. The scaling size parameter $\rho_0$ is given by 
\begin{equation}
    \rho_0=\sqrt{\frac{1}{m^*M}\sum_{i<k}m_im_kR_{ik}^2}\;,
\end{equation}
with $R_{ik}$ the two-body scaling length of the $i-k$ system, which is calculated as the outer
classical turning point for the $i-k$ potential. We choose $R_{ik}$ equal to zero for repulsive
potentials. The condition for three- and four-body quantum halos is now
$\langle \rho^2_{HR} \rangle / \rho_0^2>2$.

The dependence of the scaled size on the scaled energy for ABB and AABB are 
shown in Fig.~\ref{Fig:HaloStates}(f). We find a non-monotonic behavior, in 
clear contrast with the dependence observed in the dimer case (see 
Fig.~\ref{Fig:HaloStates}(e)).
That is, the cluster size decreases with increasing energy and reaches a 
minimum and then it starts to grow again. The minima correspond to the deepest 
bound states~\cite{PhysRevA.101.041602}. This resurgence appears as the 
clusters approach to the thresholds, where trimers eventually break into a dimer 
and an atom, and tetramers into two dimers. We want to emphasize that all the 
trimers and tetramers analyzed in Fig.~\ref{Fig:HaloStates} are halo states, 
although they are organized in significantly different spatial structures. On 
the left side of the minima, the clusters are almost radially symmetric and all 
the interparticle distances are of the same order. However, at the minima and on 
the right side of the minima the cluster structures are elongated and highly 
asymmetric.
The AABBB pentamer and AAABBB hexamer are self-bound and are manifestly halo 
states. Their mean square size has a similar behavior to the one observed before 
for the trimer and tetramer, that is a minimum corresponding to the larger 
binding energy which separates a regime of nearly symmetric particle 
distribution form another one, more elongated, and thus asymmetric.
It is important to notice that the existence of halo tetramers, pentamers and hexamers
is very unusual as it contradicts the usual tendency of self-bound clusters to shrink
and lose the halo character as the number of particles is increased. As we demonstrate
in the present work, the bilayer geometry is very promising for creation of halo states
with up to six particles which to the best of our knowledge has never been observed before.
    
A possible experimental implementation for observing the predicted halo states can be
realized by using bosonic dipolar molecules produced with mixtures of
$^{87}$Rb$^{133}$Cs~\cite{PhysRevLett.113.205301,PhysRevLett.113.255301} and
$^{23}$Na$^{87}$Rb~\cite{PhysRevLett.116.205303,PhysRevA.97.020501} characterized by
dipolar lengths $r_0\sim5\times 10^{-6}$m and $2\times10^{-5}$m, respectively. The
interlayer distance, half of the laser wavelength $\lambda$, has typical values of
$h\approx (2-5)\times 10^{-7}$m, and typical lengths of the transverse confinement
$a_\perp=(\lambda/2\pi)s^{-1/4}$ are $a_\perp\approx(3-8)\times 10^{-8}$m, where
$s\sim16$ is the potential depth of the transverse confinement in units of the recoil energy.

A possible issue of concern may be the validity of our findings in an experimental
realization, where the bilayer has a quasi-two-dimensional geometry and
not strictly two-dimensional one as in our model. To this aim, we have compared
the dimer energy of our model with the dimer energy of a quasi 2D model. For typical
experimental parameters, we have found that the change in the dimer energy is at
most 20$\%$. Therefore, it can be concluded that the effects of considering a
quasi-two dimensional model do not change our main findings.

\section{Conclusions}
We used the diffusion Monte Carlo method to study the ground-state properties of
few-body bound states of dipolar bosons in a two-dimensional bilayer setup. 
We have studied clusters composed by up to six particles, for different values of
the interlayer distance, as candidates for quantum halo states. 
In the case of dimers, we find that for values of the interlayer separation larger
than $h/r_0 = 0.45$ the clusters are halo states and they 
follow a universal scaling law. In the cases of trimers up to hexamers, we find 
two very different halo structures. For large values of the interlayer 
separation the halo structures are almost radially symmetric and the distances 
between dipoles are all of the same scale. In contrast, in the vicinity of the 
threshold for unbinding, the clusters are elongated and highly anisotropic. 
Importantly, our results prove the existence of stable halo states composed of 
up to six particles. To the best of our knowledge, this is the first time that 
halo states with such a large number of particles are predicted in 
a numerical simulation. Indeed, the addition of particles to a two or 
three body halo states typically makes them shrink towards a more compact liquid 
structure. We conclude that the bilayer geometry is advantageous for the observation of
halo states in future experiments. We hope that these results will stimulate
experimental activity in this setup, composed by atoms with dominant dipolar
interaction, to bring evidence of these quantum halo states.

In outlook, our results can stimulate further theoretical and experimental research
of halo states in ultracold gases. It could be interesting to understand how the
possibility of tunneling between the layers affects the stability of the halo states.
An interesting new path of research could be to study halo states in a bilayer system
of fermionic dipoles~\cite{PhysRevLett.111.220405,PhysRevA.90.053620}. 

\section{Method}
To investigate the structural properties of the dipolar clusters, we use a
second-order DMC method~\cite{BoronatCasulleras1994} with pure
estimators~\cite{CasullerasBoronat1995}. This method allows for an exact
estimation of the ground-state energy, as well as other properties, within
controllable statistical errors. The DMC stochastically solves the
imaginary-time Schr\"odinger equation, 
\begin{equation}
    -\frac{\partial\Psi(\boldsymbol{\rho},\tau)}{\partial\tau}=(H-E_s)
    \Psi(\boldsymbol{\rho},\tau)\;,
    \label{eq02}
\end{equation}
where $\tau=it/\hbar$ is a imaginary time, $E_s$ is an energy shift and the 
walker $\boldsymbol{\rho}=(\boldsymbol{\rho}_1,\dots,\boldsymbol{\rho}_N)$ is a vector
containing positions of $N$ particles. Importance sampling is used to reduce the
statistical noise of the calculation, which consists on rewritten the 
Schr\"odinger equation, Eq.~(\ref{eq02}), for the mixed distribution 
$\Phi(\boldsymbol{\rho},\tau)=\Psi(\boldsymbol{\rho},\tau)\psi(\boldsymbol{\rho})$.
We use a trial wave function $\psi$ of the form
\begin{equation}
\begin{aligned}
\label{eq03}
\psi(\boldsymbol{\rho}_1,\dots,\boldsymbol{\rho}_N)=&\prod_{i<j}^{N_\1}f_{\1\1}
(\rho_{ij})\prod_{\alpha<\beta}^{N_\2}f_{\2\2}(\rho_{\alpha\beta})\\\times
&\Bigg[\prod_{i=1}^{N_\1}\sum_{\alpha=1}^{N_\2}f_{\1\2}(\rho_{i\alpha})+
\prod_{\alpha=1}^{N_\2}\sum_{i=1}^{N_\1}f_{\1\2}(\rho_{i\alpha})\Bigg]\;,
\end{aligned}
\end{equation}
which is suitable for describing systems with short-range correlations and as well as
long-range asymptotic behavior.

The trial wave function for intraspecies correlations is built from the
zero-energy two-body scattering solution 
\begin{equation}
f_{\1\1}(\rho)=f_{\2\2}(\rho)=C_0K_0(2\sqrt{r_0/\rho})\;,
\label{eq04}
\end{equation}
$K_0(\rho)$ being the modified Bessel function and $C_0$ a constant. The interspecies
interactions are described by the dimer wave function $f_{\1\2}(\rho)$ up to $R_0$.
From the variational distance $R_0$ on we took the free scattering solution
$f_{\1\2}(\rho)=CK_0(\sqrt{-mE_{\1\2}}\rho/\hbar)$. We impose continuity of the
logarithmic derivative at the matching point $R_0$, yielding the following equality
\begin{equation}
\frac{ f^{'}_{\1\2}(R_0)}{f_{\1\2}(R_0)}=-\frac{\sqrt{-mE_{\1\2}}}{\hbar} 
\frac{K_1(\sqrt{-mE_{\1\2}}R_0/\hbar)}{K_0(\sqrt{-mE_{\1\2}}R_0/\hbar)}\;.  
\label{eq05}
\end{equation} 

In a DMC calculation, the expectation value of an observable $\hat{O}$ is obtained
for long enough imaginary time propagation
\begin{equation}
\langle\hat{O}\rangle_{mix}=\frac{\langle\psi|\hat{O}|\Psi_0\rangle}
{\langle\psi|\Psi_0\rangle}=\lim_{\tau\to\infty}\frac{\int 
d\boldsymbol{\rho}\psi(\boldsymbol{\rho}) \hat{O} \Psi(\boldsymbol{\rho},\tau) }
{\int d\boldsymbol{\rho}\psi(\boldsymbol{\rho}) \Psi(\boldsymbol{\rho},\tau)}\;,
\label{eq06}
\end{equation}
with $\Psi_0$ the ground-state wave function. The last equation is known as the
\textit{mixed estimator}. Equation~(\ref{eq06}) gives the exact expectation value
for the Hamiltonian and for observables that commute with it. In the case of
operators that do not commute with $\hat{H}$, the result obtained from
Eq.~(\ref{eq06}) can be biased by $\psi$. In this case, it is possible to obtain
exact expectation values using the pure estimators technique~\cite{CasullerasBoronat1995}.
In the present study, pure estimators are used for the calculation of the pair distribution 
functions, the spatial distribution functions, and the size of the clusters. 

\section{Acknowledgments}
This work has been supported by the Ministerio de Economia, Industria y Competitividad
(MINECO, Spain) under grant FIS2017-84114-C2-1-P. We also acknowledge financial support
from Secretaria d'Universitats i Recerca del Departament d'Empresa i Coneixement de la
Generalitat de Catalunya, co-funded by the European Union Regional Development Fund
within the ERDF Operational Program of Catalunya (project QuantumCat, ref.~001-P-001644).
The authors thankfully acknowledge the computer resources at
Cibeles and the technical support provided by Barcelona
Supercomputing Center (RES-FI-2020-2-0020). G.G. acknowledges a fellowship from CONACYT (M\'exico).

\bibliography{bibliography} 

\begin{thebibliography}{10}

\bibitem{EFIMOV1970563}
V.~Efimov, ``Energy levels arising from resonant two-body forces in a
  three-body system,'' {\em Physics Letters B}, vol.~33, no.~8, pp.~563 -- 564,
  1970.

\bibitem{kraemer2006evidence}
T.~Kraemer, M.~Mark, P.~Waldburger, J.~G. Danzl, C.~Chin, B.~Engeser, A.~D.
  Lange, K.~Pilch, A.~Jaakkola, H.-C. N{\"a}gerl, {\em et~al.}, ``Evidence for
  {Efimov} quantum states in an ultracold gas of caesium atoms,'' {\em Nature},
  vol.~440, no.~7082, pp.~315--318, 2006.

\bibitem{Naidon_2017}
P.~Naidon and S.~Endo, ``Efimov physics: a review,'' {\em Reports on Progress
  in Physics}, vol.~80, p.~056001, mar 2017.

\bibitem{sachdev_2009}
S.~Sachdev, {\em Quantum phase transitions}.
\newblock Cambridge Univ. Press, 2009.

\bibitem{Bloch2005}
I.~Bloch, ``Ultracold quantum gases in optical lattices,'' {\em Nat. Phys.},
  vol.~1, pp.~23--30, 10 2005.

\bibitem{ReviewFermiGases}
S.~Giorgini, L.~P. Pitaevskii, and S.~Stringari, ``Theory of ultracold atomic
  {Fermi} gases,'' {\em Rev. Mod. Phys.}, vol.~80, pp.~1215--1274, Oct 2008.

\bibitem{Zwerger2012book}
W.~Zwerger, {\em The BCS-BEC crossover and the unitary Fermi gas}.
\newblock Heidelberg New York: Springer, 2012.

\bibitem{RevModPhys.66.1105}
K.~Riisager, ``Nuclear halo states,'' {\em Rev. Mod. Phys.}, vol.~66,
  pp.~1105--1116, Jul 1994.

\bibitem{TANIHATA1985}
I.~Tanihata, H.~Hamagaki, O.~Hashimoto, S.~Nagamiya, Y.~Shida, N.~Yoshikawa,
  O.~Yamakawa, K.~Sugimoto, T.~Kobayashi, D.~Greiner, N.~Takahashi, and
  Y.~Nojiri, ``Measurements of interaction cross sections and radii of {He}
  isotopes,'' {\em Physics Letters B}, vol.~160, no.~6, pp.~380 -- 384, 1985.

\bibitem{PhysRevLett.55.2676}
I.~Tanihata, H.~Hamagaki, O.~Hashimoto, Y.~Shida, N.~Yoshikawa, K.~Sugimoto,
  O.~Yamakawa, T.~Kobayashi, and N.~Takahashi, ``Measurements of interaction
  cross sections and nuclear radii in the light $p$-shell region,'' {\em Phys.
  Rev. Lett.}, vol.~55, pp.~2676--2679, Dec 1985.

\bibitem{RevModPhys.82.1225}
C.~Chin, R.~Grimm, P.~Julienne, and E.~Tiesinga, ``Feshbach resonances in
  ultracold gases,'' {\em Rev. Mod. Phys.}, vol.~82, pp.~1225--1286, Apr 2010.

\bibitem{Jensen2004}
A.~S. Jensen, K.~Riisager, D.~V. Fedorov, and E.~Garrido, ``Structure and
  reactions of quantum halos,'' {\em Rev. Mod. Phys.}, vol.~76, pp.~215--261,
  Feb 2004.

\bibitem{Riisager_2013}
K.~Riisager, ``Halos and related structures,'' {\em Physica Scripta},
  vol.~T152, p.~014001, jan 2013.

\bibitem{RIISAGER1992393}
K.~Riisager, A.~Jensen, and P.~MÃžller, ``Two-body halos,'' {\em Nuclear
  Physics A}, vol.~548, no.~3, pp.~393 -- 413, 1992.

\bibitem{PhysRevC.49.201}
D.~V. Fedorov, A.~S. Jensen, and K.~Riisager, ``Three-body halos: Gross
  properties,'' {\em Phys. Rev. C}, vol.~49, pp.~201--212, Jan 1994.

\bibitem{PhysRevLett.113.253401}
P.~Stipanovi{\'c}, L.~Vranje{\v{s}}~Marki{\'c}, I.~Be{\v{s}}li{\'c}, and
  J.~Boronat, ``Universality in molecular halo clusters,'' {\em Phys. Rev.
  Lett.}, vol.~113, p.~253401, Dec 2014.

\bibitem{Stipanovi2017QuantumHS}
P.~Stipanovi{\'c}, L.~V. Marki{\'c}, and J.~Boronat, ``Quantum halo states in
  helium tetramers.,'' {\em The journal of physical chemistry. A}, vol.~121 1,
  pp.~308--314, 2017.

\bibitem{Stipanovic2019}
P.~Stipanović, L.~Vranje{\v{s}}~Marki{\'c}, A.~Gudyma, and J.~Boronat,
  ``Universality of size-energy ratio in four-body systems,'' {\em Scientific
  Reports}, vol.~9, 12 2019.

\bibitem{PhysRevLett.103.033004}
J.~P. D'Incao, J.~von Stecher, and C.~H. Greene, ``Universal four-boson states
  in ultracold molecular gases: Resonant effects in dimer-dimer collisions,''
  {\em Phys. Rev. Lett.}, vol.~103, p.~033004, Jul 2009.

\bibitem{PhysRevA.90.043631}
X.~Y. Yin, D.~Blume, P.~R. Johnson, and E.~Tiesinga, ``Universal and
  nonuniversal effective $n$-body interactions for ultracold harmonically
  trapped few-atom systems,'' {\em Phys. Rev. A}, vol.~90, p.~043631, Oct 2014.

\bibitem{Kievsky_2014}
A.~Kievsky, M.~Gattobigio, and E.~Garrido, ``Universality in few-body systems:
  from few-atoms to few-nucleons,'' {\em Journal of Physics: Conference
  Series}, vol.~527, p.~012001, jul 2014.

\bibitem{Nielsen1997}
E.~Nielsen, D.~V. Fedorov, and A.~S. Jensen, ``Three-body halos in two
  dimensions,'' {\em Phys. Rev. A}, vol.~56, p.~3287, 1997.

\bibitem{Nielsen1999}
E.~Nielsen, D.~V. Fedorov, and A.~S. Jensen, ``Structure and occurrence of
  three-body halos in two dimensions,'' {\em Few-Body Syst.}, vol.~27, p.~15,
  1999.

\bibitem{LandauLifshitz_iii}
L.~D. Landau and E.~M. Lifshitz, {\em Quantum Mechanics : Non-Relativistic
  Theory}.
\newblock Burlington: Elsevier Science, 1977.

\bibitem{Yudson1997}
V.~I. Yudson, M.~G. Rozman, and P.~Reineker, ``Bound states of two particles
  confined to parallel two-dimensional layers and interacting via dipole-dipole
  or dipole-charge laws,'' {\em Phys. Rev. B}, vol.~55, p.~5214, Feb 1997.

\bibitem{Shih2009}
S.-M. Shih and D.-W. Wang, ``Pseudopotential of an interaction with a power-law
  decay in two-dimensional systems,'' {\em Phys. Rev. A}, vol.~79, p.~065603,
  2009.

\bibitem{Armstrong2010}
J.~R. Armstrong, N.~T. Zinner, D.~V. Fedorov, and A.~S. Jensen, ``Bound states
  and universality in layers of cold polar molecules,'' {\em Europhys. Lett.},
  vol.~91, p.~16001, 2010.

\bibitem{Klawunn2010}
M.~Klawunn, A.~Pikovski, and L.~Santos, ``Two-dimensional scattering and bound
  states of polar molecules in bilayers,'' {\em Phys. Rev. A}, vol.~82,
  p.~044701, Oct 2010.

\bibitem{Baranov2011}
M.~A. Baranov, A.~Micheli, S.~Ronen, and P.~Zoller, ``Bilayer superfluidity of
  fermionic polar molecules: Many-body effects,'' {\em Phys. Rev. A}, vol.~83,
  p.~043602, Apr 2011.

\bibitem{Volosniev2011}
A.~G. Volosniev, D.~V. Fedorov, A.~S. Jensen, and N.~T. Zinner, ``Model
  independence in two dimensions and polarized cold dipolar molecules,'' {\em
  Phys. Rev. Lett.}, vol.~106, p.~250401, 2011.

\bibitem{Macia2014}
A.~Macia, G.~E. Astrakharchik, F.~Mazzanti, S.~Giorgini, and J.~Boronat,
  ``Single-particle versus pair superfluidity in a bilayer system of dipolar
  bosons,'' {\em Phys. Rev. A}, vol.~90, p.~043623, Oct 2014.

\bibitem{Simon1976}
B.~Simon, ``The bound state of weakly coupled {Schr\"{o}dinger} operators in
  one and two dimensions,'' {\em Annals of Physics}, vol.~97, p.~279, Apr.
  1976.

\bibitem{Volosniev2012}
A.~G. Volosniev, D.~V. Fedorov, A.~S. Jensen, and N.~T. Zinner, ``Few-body
  bound-state stability of dipolar molecules in two dimensions,'' {\em Phys.
  Rev. A}, vol.~85, p.~023609, Feb 2012.

\bibitem{PhysRevA.101.041602}
G.~Guijarro, G.~E. Astrakharchik, J.~Boronat, B.~Bazak, and D.~S. Petrov,
  ``Few-body bound states of two-dimensional bosons,'' {\em Phys. Rev. A},
  vol.~101, p.~041602, Apr 2020.

\bibitem{BoronatCasulleras1994}
J.~Boronat and J.~Casulleras, ``{Monte Carlo} analysis of an interatomic
  potential for {He},'' {\em Phys. Rev. B}, vol.~49, pp.~8920--8930, Apr 1994.

\bibitem{CasullerasBoronat1995}
J.~Casulleras and J.~Boronat, ``Unbiased estimators in quantum {Monte Carlo}
  methods: Application to liquid $^{4}\mathrm{He}$,'' {\em Phys. Rev. B},
  vol.~52, pp.~3654--3661, Aug 1995.

\bibitem{PhysRevLett.113.205301}
T.~Takekoshi, L.~Reichs\"ollner, A.~Schindewolf, J.~M. Hutson, C.~R. Le~Sueur,
  O.~Dulieu, F.~Ferlaino, R.~Grimm, and H.-C. N\"agerl, ``Ultracold dense
  samples of dipolar {RbCs} molecules in the rovibrational and hyperfine ground
  state,'' {\em Phys. Rev. Lett.}, vol.~113, p.~205301, Nov 2014.

\bibitem{PhysRevLett.113.255301}
P.~K. Molony, P.~D. Gregory, Z.~Ji, B.~Lu, M.~P. K\"oppinger, C.~R. Le~Sueur,
  C.~L. Blackley, J.~M. Hutson, and S.~L. Cornish, ``Creation of ultracold
  $^{87}\mathrm{Rb}^{133}\mathrm{Cs}$ molecules in the rovibrational ground
  state,'' {\em Phys. Rev. Lett.}, vol.~113, p.~255301, Dec 2014.

\bibitem{PhysRevLett.116.205303}
M.~Guo, B.~Zhu, B.~Lu, X.~Ye, F.~Wang, R.~Vexiau, N.~Bouloufa-Maafa,
  G.~Qu\'em\'ener, O.~Dulieu, and D.~Wang, ``Creation of an ultracold gas of
  ground-state dipolar $^{23}\mathrm{Na}^{87}\mathrm{Rb}$ molecules,'' {\em
  Phys. Rev. Lett.}, vol.~116, p.~205303, May 2016.

\bibitem{PhysRevA.97.020501}
M.~Guo, X.~Ye, J.~He, G.~Qu\'em\'ener, and D.~Wang, ``High-resolution internal
  state control of ultracold $^{23}\mathrm{Na}^{87}\mathrm{Rb}$ molecules,''
  {\em Phys. Rev. A}, vol.~97, p.~020501(R), Feb 2018.

\bibitem{PhysRevLett.111.220405}
N.~Matveeva and S.~Giorgini, ``Impurity problem in a bilayer system of
  dipoles,'' {\em Phys. Rev. Lett.}, vol.~111, p.~220405, Nov 2013.

\bibitem{PhysRevA.90.053620}
N.~Matveeva and S.~Giorgini, ``Fixed-node diffusion monte carlo study of the
  bcs-bec crossover in a bilayer system of fermionic dipoles,'' {\em Phys. Rev.
  A}, vol.~90, p.~053620, Nov 2014.

\end{thebibliography}
\bibliographystyle{ieeetr}

\end{document}